\documentclass[a4paper,twoside,UTF-8]{article}
\newcommand{\ucite}[1]{$^{[#1]}$}
\usepackage{AMS2LA,fancyhdr,xcolor,multicol,graphics,bm,cmap,graphicx, CJKutf8}
\usepackage{amssymb,gensymb,mathcomp}
\usepackage[pdfstartview=FitH, bookmarks=false, colorlinks=false,
    linkcolor=blue, urlcolor=blue, pdfborder=001, citecolor=blue,
    pdftitle={Changepdftitle}, pdfauthor={Changepdfauthor},
    pdfcreator=CHIN. PHYS. LETT., pdfproducer=CPL,
    pdfkeywords={changePACS}]{hyperref}
\graphicspath{{figs2/}}

\newcommand{\explel}{{\large\sffamily~~~~~~~~~~~~~~~~~} \hfill}
\newcommand{\expler}{\hfill\fcolorbox{lightgray}{lightgray}{\textcolor{red}{{\large\fontfamily{phv}\textbf{New Submission}}}}}

\def\headrule{\kern 1mm \hrule width 17cm \kern -1mm}%
\def\footnoterule{\kern 1mm \hrule width 7cm \kern 2.2mm}%
\def\REF#1{\par\hangindent\parindent\indent\llap{#1\enspace}\ignorespaces}%

\renewcommand{\cite}[1]{\,[#1]}
\newcommand{\etal}{\textit{et~al}.\ }
\newcommand{\cplyear}{2022}
\newcommand{\cplvol}{x}
\newcommand{\cplno}{x}
\newcommand{\cplpagenumber}{xx{xxxx}}
\newcommand{\cplpage}{\cplpagenumber-\thepage}

\begin{document}
\vspace* {-6mm}
\begin{center}
\large\bf{\boldmath{Efficient two-dimensional defect-free dual-species atom arrays rearrangement algorithm with near-fewest atom moves}}
\footnotetext{\hspace*{-5.4mm}Z.T. and L.Y contributed equally to this work. Supported by the National Key R\&D Program of China (Grant No. 2021YFA1402001, No. 2017YFA0304501), the Youth Innovation Promotion Association CAS No. Y2021091 and No. 2019325, the National Natural Science Foundation of China under Grant No. U20A2074 and No. 12074391, the K.C. Wong Education Foundation (Grant No. GJTD-2019-15).

\noindent$^{*}$Corresponding author. Email: xupeng@apm.ac.cn

\noindent\copyright\,{\cplyear}
\href{http://www.cps-net.org.cn}{Chinese Physical Society} and
\href{http://www.iop.org}{IOP Publishing Ltd}}
\\[6mm]
\normalsize \rm{}
\begin{CJK}{UTF8}{gbsn}
	Zhi-Jin Tao(陶知进)$^{1,2}$, Li-Geng Yu(余立庚)$^{1,2}$, Peng-Xu(许鹏)$^{1,3*}$, Jia-Yi Hou(侯嘉毅)$^{1}$, Xiao-Dong He(何晓东)$^{1}$ and Ming-Sheng Zhan(詹明生)$^{1,3}$
\end{CJK}
\\[2mm]\small\sl

$^{1}$State Key Laboratory of Magnetic Resonance and Atomic and Molecular Physics, Wuhan Institute of Physics and Mathematics, Innovation Academy for Precision Measurement Science and Technology, Chinese Academy of Sciences, Wuhan 430071, China

$^{2}$School of Physics and Technology, Wuhan University, Wuhan 430072, China

$^{3}$Wuhan Institute of Quantum Technology, Wuhan 430206, China
\\[4mm]\normalsize\rm{}(Received xx xx 2022; accepted xxx; published online )
\end{center}
\vskip 1.5mm

\noindent{
	\narrower\small{}
	Dual-species single-atom array in optical tweezers has several advantages over the single-species atom array as a platform for quantum computing and quantum simulation. Thus, creating the defect-free dual-species single-atom array with atom numbers over hundreds is essential. As recent experiments demonstrated, one of the main difficulties lies in designing an efficient algorithm to rearrange the stochastically loaded dual-species atoms arrays into arbitrary demanded configurations. We propose a heuristic connectivity optimization algorithm (HCOA) to provide the near-fewest number of atom moves. Our algorithm introduces the concept of using articulation points in an undirected graph to optimize connectivity as a critical consideration for arranging the atom moving paths. Tested in array size of hundreds atoms and various configurations, our algorithm shows a high success rate ($> 97\%$), low extra atom moves ratio, good scalability, and flexibility. Furthermore, we proposed a complementary step to solve the problem of atom loss during the rearrangement. \par}
\vskip 3mm
\normalsize\noindent{\narrower{PACS: 03.67.Lx, 87.55.kd, 37.10.Gh}

{\rm\hspace*{13mm}DOI: 10.1088/****-****/\cplvol/\cplno/\cplpagenumber}

\par}\vskip 6mm
\begin{multicols}{2}

In the physical implementation of quantum computation, neutral atom array has become a promising candidate for its various and controllable atom-atom interactions (like the Van der Wass interaction and Rydberg-Rydberg interaction) and good scalability\ucite{1-4}. With the demonstration of high fidelity single qubit gates\ucite{5}, entangled states preparation and measurement\ucite{5-8}, as well as the CNOT gate demonstration\ucite{9}, neutral atom array system has come to a stage of realizing quantum error correction(QEC).
In this phase, the mixed-species qubit system is more advantageous than the single species qubit setup with the requirements for simultaneously gate operating and state detecting\ucite{10,11}.
Especially, the checker board dual-atom configuration in a square optical lattice is a promising platform for implementing large scale quantum error correction based on surface code, because of the reduced cross-talk when implementing the quantum gate and measuring, due to the large resonant frequency difference between the different kinds of atoms\ucite{12}.

In the field of quantum simulation, Rydberg atom array serves as an excellent platform to simulate the quantum dynamic model and optimization problems mapped into Ising model and so on by controlling the interactions between atoms\ucite{13, 14}. With the dual-species system realized, more types of interactions can be realized, which provides the feasibility to investigate new phenomenons. Also, building single molecules by merging two single atoms has been achieved by different scientific groups recently\ucite{15-18}. Thus, the dual-species atom array can also be utilized to construct an atom-molecule system by manipulating two kinds of atoms.

To bring the proposed schemes into practice, preparing an arbitrary user-defined atom array composed of multi-species atoms is indispensable. In the experiment, one of the feasible ways is to prepare many optical traps nearly half fulfilled after initial loading, then rearrange single atoms one by one using movable tweezers to achieve the desired atom array configuration\ucite{19}. In this protocol, one of the most intractable obstacles is to develop an algorithm to output an optimized moving path when inputting an initial atom configuration. In the case of single species, there are several efficient and scalable algorithms, including HHA, HPFA, and Hungarian matching\ucite{20-23}. As a result, the single-species atom array has been scaled to hundreds of atoms and from one dimension to three dimensions\ucite{24-27}.
For the defect-free multi-species atom array, there are mainly two experimental routes. One is to build two setups of different wavelengths for separate species, including trap arrays and moving tweezers. In this way, atoms of different species will be stochastically loaded into the traps of corresponding wavelength\ucite{28,29}. This route can directly use the extension of single species algorithms\ucite{20-23} for many geometries but suffers from more complicated experimental setups, thus hasn't demonstrated defect-free atom arrays.
The other way is to use the identical trap array and moving tweezer for two species. Using the 808-nm tweezer array, C.Sheng \etal have realized defect-free two dimensional 6 $\times$ 4 dual-species atom assembly of $^{85}$Rb ($^{87}$Rb) atoms with a filling fraction of 0.88 (0.89)\ucite{19}.
However, due to the more complicated situation that the target sites may be filled with unmatched atoms and different types of atoms will block each other in the moving path, the preliminary algorithm demonstrated in C.Sheng \etal's work\ucite{19} is facing a limitation that the success probability decays exponentially with the number of atoms increasing.

To provide an efficient, well-optimized, and widely applicable algorithm for mixed-species atom rearrangement, here we propose heuristic connectivity optimized algorithm (HCOA). This algorithm can solve the limitation of the success rate in the second route, and can also be applied to the first route. HCOA is based on the undirected graph which is mapped from dual-species atom distributions. By taking the connectivity of the graph as a critical consideration at the main steps, the success rate of HCOA decreases linear with the number of atoms increasing and remains larger than 0.80 as the number of target atoms increase to 3600. To trap more atoms in actual experimental conditions, the algorithm set that an atom can only be transported along the lattice links\ucite{15}. Moreover, recent experimental researches\ucite{14, 15} have imposed several constraints. First, the algorithm should run in milliseconds magnitude for hundreds of atoms to minimize the total assembly time. Second, the algorithm should reduce the number of atom moves firstly and then shorten the total transport path, because capturing and releasing an atom at the beginning and the end of transport process costs relatively longer time and has limited success rate. Thus, we should reduce the number of atom moves in priority to reach a high experimental success rate.

\vskip 4mm

\centerline{\includegraphics{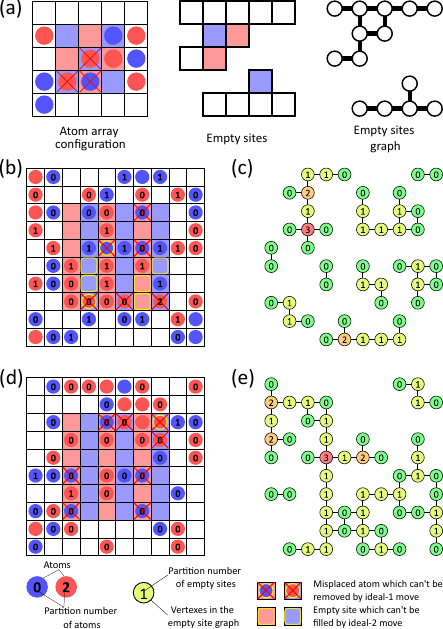}}

\vskip 2mm

\centerline{\footnotesize \begin{tabular}{p{7.5cm}}\bf Fig.\,1. \rm
		Examples of dual-species atom arrays and the corresponding empty-site graphs . For schemes of atom arrays, the red (blue) disks denote the single atoms of $^{87}$Rb ($^{85}$Rb), the red(blue)-colored blocks denote the target site of $^{87}$Rb ($^{85}$Rb). For schemes of corresponding empty site graphs, circles denote vertexes, and lines denote edges. Numbers on atoms and empty sites denote the partition number (non-movable atoms have no partition number).(a) Scheme of extracting empty site graphs from atom arrays. (b) An example of atom array after initial stochastic loading. (c) Corresponding empty site graph for (a). (d) Another example of dual-species atom array, whose number of isolated empty site regions is greatly reduced. (e) Corresponding empty site graph for (c).
\end{tabular}}

\vskip 0.5\baselineskip

\vskip 4mm
\end{multicols}
\centerline{\includegraphics{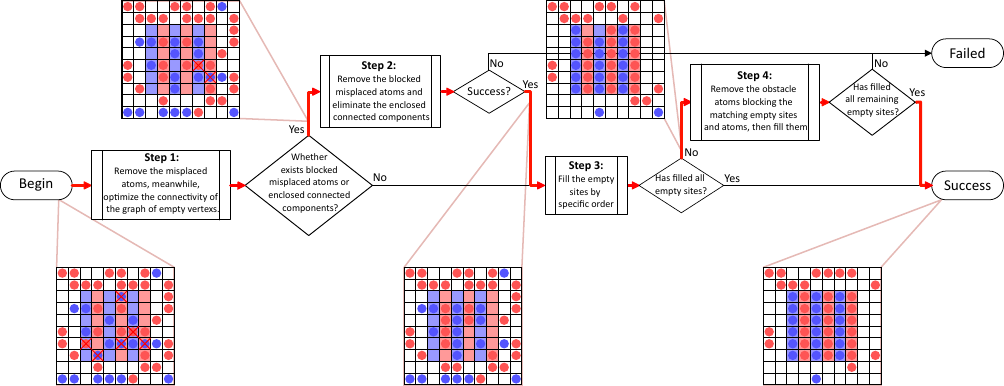}}

\vskip 2mm

\centerline{\footnotesize \begin{tabular}{p{17cm}}\bf Fig.\,2. \rm
	The brief flowchart of HCOA, with a example that how HCOA process a stochastically loaded atom array.
\end{tabular}}

\vskip 0.5\baselineskip

\begin{multicols}{2}
We start the analysis from one single atom move to approach the near-fewest atom moves. Each single atom move can be decomposed into two parts. The first part is to choose one atom to move, while the second part is to choose an empty site for the chosen atom. After initial stochastic loading, three types of atom-site pairs appear in the atom array. They are misplaced atoms in the wrong sites, atoms in the suitable sites, and outside atoms which do not lie in the target region. As for the empty sites, we classify them into two types, the empty target sites, and the empty outside sites. Totally, there are six types of atom moves according to the classification.
Only two types are not redundant and even necessary to achieve the fewest atom moves. One type is moving a misplaced atom to a matched empty target site,  which can decrease the number of misplaced atoms and empty target sites. The other type is moving an outside atom to a target empty site to reduce the number of empty target sites. We call the two types of moves ideal moves and refer to the first type as ideal-1 move while the latter as ideal-2 move.
	
After initial stochastic loading, like Fig.1 (b), there are a lot of isolated connected empty site regions, which makes empty sites have fewer atoms to choose from, same as misplaced atoms. On the contrary, the number of ideal moves will increase if the number of isolated, empty site regions becomes less, as shown in Fig.1 (d). Therefore, we can try to merge the isolated, empty site regions as much as possible to improve the connectivity of entire empty site regions in each atom move.


We abstract all the empty sites as an undirected graph $G<V, E>$. The vertex set $V$ consists of all empty sites. As for the edge set $E$, for empty sites $i,j\in V$, if an atom can be transported from site $i$ to $j$ safely, then the tuple $(i,j)\in E$. And the isolated, empty site region is equivalent to the concept of the connected component in graph theory. See Fig.1(a) for a simple example of mapping the empty sites to an undirected graph. To analyze how the number of connected components changes, a new concept called partition number is introduced. The partition number of an atom $a$ for a graph $G$, denoted as $P(G,a)$, is defined as the reduction number of connected components of the graph $G$ if we treat the atom which is movable as an empty site and add the empty site and its connected edges to the empty site graph $G$. Furthermore, non-movable atoms have no partition number. The partition number of an empty site $s$ for a graph $G$, denoted as $P(G,s)$ is defined as the increment number of the connected components if the empty site $s$ and its connected edges are deleted from the graph $G$. Fig.1(b-d) illustrates the concept of the partition number. From the definition, in each atom move, to reduce the connected components and improve the graph's connectivity, we can consider as follows. In choosing which atom to move, we can choose the atom with a larger partition number and then update the graph. In choosing empty sites to move the chosen atom into, we should choose the empty sites with the minimum partition number. Then, the graph's connectivity will get improved after such a move. In the supplementary materials, we introduced the concept of {\it branch factor} to prove the effectiveness of improving the connectivity of the graph.

Based on the above analysis, the general steps of our algorithm are as follows: (1) Move the misplaced atoms to the appropriate empty sites directly (i.e., do the ideal-1 moves as more as possible), and the partition number will be taken as an essential consideration. (2) Remove the misplaced atoms left unsolved
\end{multicols}

\centerline{\includegraphics{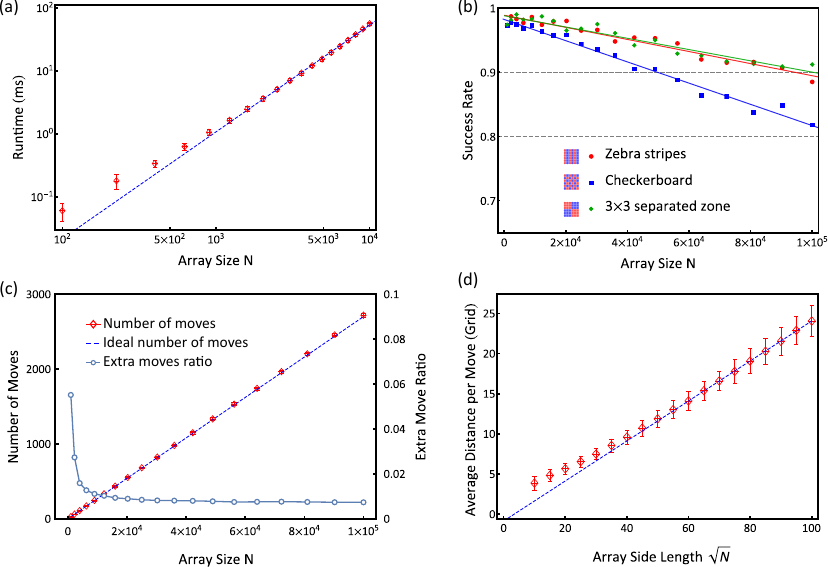}}

\vskip 2mm
\centerline{\footnotesize \begin{tabular}{p{14cm}}\bf Fig.\,3. \rm
	Simulation results. The errorbars in sub-figure (a,c-d) represent the standard deviation. (a) Runtime of HCOA versus array size (zebra stripes pattern). The dash line is the linear fitting of the last 13 data, having slope $1.70\pm0.02$, which indicates the time complexity of the algorithm is $O(N^2)$. (b) Success rate of different patterns versus array size. The success rate of HCOA algorithm is fitted with a linear curve $y=a-bx$, for Zebra strips pattern $a=0.989\pm0.003$, $b=(9.4\pm0.6)\times10^{-6}$, for Checkerboard pattern $a=0.982\pm0.003$, $b=(16.5\pm0.5)\times10^{-6}$ and for $3\times3$ separated zone $a=0.988\pm0.003$, $b=(8.8\pm0.6)\times10^{-6}$. (c) Number of moves, ideal number of moves and extra move ratio versus array size (zebra strips pattern). (d) Average distance per move versus array side length (zebra stripes pattern). The dash line is the linear fitting of the last 9 datas.
\end{tabular}}

\vskip 0.5\baselineskip

\begin{multicols}{2}
\noindent by step 1 and let all target empty site regions connected with the outside by moving matched atoms in the target region. (3) Fill the empty target sites with the matched outside atoms (i.e., make the ideal-2 moves). (4) Solve the most common unsolved cases left by step 3. The brief flowchart is in Fig.2. The details of each step are in the supplementary materials.
	
Our algorithm is simulated in the array of hundreds of atoms and different configurations, are shown in Fig.3. The runtime data is performed with a C++ code with gcc compiler opening -O3 optimization and an AMD CPU Ryzen 7 5700U. Three target array pattern ``Zebra Stripes'', ``Checkerboard'' or ``$3\times3$ separated zone'' are respectively tessellation of $1\times N_T^{\frac{1}{2}}$, $1\times 1$ or $3\times 3$ block of the two different species atoms (as demonstrated in \ucite{19}). The target region with the size of $N_T/N=0.36$, is located at the center of the trap array, where $N_T$ is the number of target sites. Each data is tested 1000 times. Without special instruction, the loading ratio is 0.5, and the ratio between the two species is 1:1. The runtime curve (see Fig.3(a)) shows that the time complexity of the algorithm is approximate $O(N^2)$. The runtime of HCOA is less than 1 ms for intermediate-scale array ($N<700$) and around 60ms for 100$\times$100 atom array with 60$\times$60 target array, showing HCOA is efficient enough for current and near-future dual-species atom array preparation experiment. More importantly, the success rate of HCOA decreases
linear with the number of atoms increasing [see Fig.3(b)] which is high ($>0.97$) for intermediate-scale arrays and remains at a high level ($>0.8$) as array size increases.
The mean number of atom moves increases linearly while the extra moves ratio decays as the array size increases. The calculation of the ideal number of atom moves is demonstrated in the supplementary materials. Fig.3(d) shows that the average move distance per move increases linearly with the array side length.

However, we find that HCOA is not good at rearranging with a high loading ratio. Taking the 10$\times$ 10 zebra strip pattern as an example, Fig.4(a) shows that the HCOA success rate decreases rapidly under a higher loading ratio because more unsolvable cases appear. After executing HCOA, a typical example of the unsolvable cases are shown in Fig.4(c), which is a near full-filled atom array with a small number of empty sites. This is very similar to the case of atom loss during the rearrangement which is caused by the finite transfer efficiency and the collision with background gas. To solve this problem, we add following complementary steps. As the connected empty site region marked as {\it connected component}, we first find out one connected component which possesses at least one target empty site adjacent to the outside region.  Second, we connect the found region with the nearest connected component through moving atoms blocking them. Third, we link the connected components which have outside empty sites with the misplaced atoms and connected components in the target region, through moving atoms [Fig.(c-d)]. Finally, we execute the former four steps (HCOA) again [Fig.4(e)]. The three procedures can loop for customized times to achieve a high success rate.  Fig.4(b) shows the effectiveness of the complementary steps.
\vskip 4mm

\centerline{\includegraphics{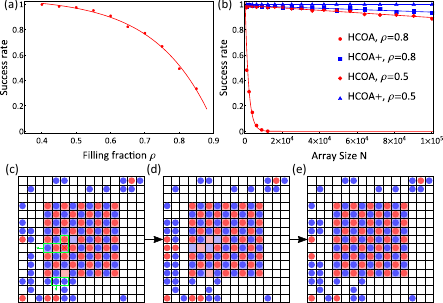}}

\vskip 2mm

\centerline{\footnotesize \begin{tabular}{p{7.5cm}}\bf Fig.\,4. \rm
		Simulation result for HCOA and HCOA+ (HCOA+ is HCOA plus the complementary step) under high loading ratio $\rho$ and scheme for the complementary step. (a) The success rate of HCOA vs loading ratio (N=100, zebra stripes pattern). (b) The success rate of HCOA and HCOA+ vs array size $N$ under loading ratios 0.5 and 0.8. (c) A typical case that HCOA cannot solve. The atoms the complementary step decides to move are highlighted. The upper highlighted atoms are moved by the first step while the lower highlighted atoms are moved by the second step. (d) The obstacle atoms are moved into outside empty sites. (e) The target region is properly filled using HCOA.\end{tabular}}

\vskip 0.5\baselineskip

In the end, we simulate the rearrangement of Einstein portrait in a 80$\times$80 square lattice, an example in a kagome lattice in Fig.5 to demonstrate the scalability of HCOA, which can be extended to rearrange almost all the atom configurations. The corresponding demonstration movies are uploaded to \url{https://github.com/YLG-WHU/Dual-species-atom-rearrangement-demonstration}.

In summary, we proposed a heuristic rearrangement algorithm for multi-species. The simulation results indicate that HCOA has a high success rate and fast speed for various multi-species target atom patterns and grid sizes ranging from 10$\times$10 to 100$\times$100. Compared with the former one in [19], whose algorithm decays exponentially, our algorithm shows a much higher success rate(see the supplementary materials for the comparison diagram). Moreover, because our algorithm is based on the undirected graph, it can be easily extended to various optical trap configurations and transport ways defined by users. In the future, it is promised to control more mobile tweezers simultaneously for rearrangement in order to reduce the preparation time. Our algorithm can also be modified to adapt the many-optical-tweezer rearrangement by simultaneously considering several objects in the priority list using parallel computing or adjusting the move order of the final output move path.

\vskip 4mm

\centerline{\includegraphics{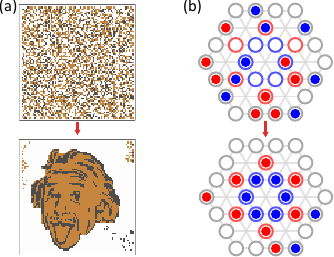}}

\vskip 2mm

\centerline{\footnotesize \begin{tabular}{p{7.5cm}}\bf Fig.\,5. \rm
Two rearrangement examples by HCOA from the stochastic initial configuration. (a) Einstein's portrait in 80$\times$80 a square lattice, where each optical trap is represented by a square block and the square blocks with different colors represent atoms of different types, respectively. Atoms can only be transported between the adjacent optical traps. (b) Kagome lattice where straight lines indicate atoms can be transported between two traps connected by them.
\end{tabular}}

\vskip 0.5\baselineskip

\section*{\Large\bf References}

\vspace*{-0.8\baselineskip}\frenchspacing

\hskip 7pt

{\footnotesize

\REF{[1]} Wilk T, Gaëtan A, Evellin C, Wolters J, Miroshnychenko Y, Grangier P and  Browaeys A 2010 {\it Phys. Rev. Lett.} {\bf 104} 010502

\REF{[2]} Henriet L, Beguin L, Signoles A, Lahaye T, BrowaeysA , Reymond G O, and Jurczak C 2020 {\it Quantum} {\bf 4} 327

\REF{[3]} Saffman M 2018 {\it Natl. Sci. Rev.} {\bf 6} 24

\REF{[4]} Saffman M, Walker T G,  and M\o{}lmer K 2010 {\it Rev. Mod. Phys.} {\bf 82} 2313

\REF{[5]} Xia T, Lichtman M, Maller K, Carr A, Piotrowicz M, Isenhower L, and Saffman M 2015 {\it Phys. Rev. Lett.} {\bf 114} 100503

\REF{[6]} Fu Z, Xu P, Sun Y, Liu Y, He X, Li X, Liu M, Li R, Wang J, Liu L, and Zhan M S 2021 {\it arXiv:2109.02491[quant-ph]}

\REF{[7]} Levine H, Keesling A, Semeghini G, Omran A, Wang T T, Ebadi S, Bernien H, Greiner M, Vuleti\ifmmode \acute{c}\else \'{c}\fi{} V, Pichler H, and Lukin M D 2019 {\it Phys. Rev. Lett.} {\bf 123} 170503

\REF{[8]} Wu T Y, Kumar A, Giraldo F, and Weiss D S 2019 {\it Nat. Phys.} {\bf 15} 538

\REF{[9]} Isenhower L, Urban E, Zhang X L, Gill A T, Henage T, Johnson T A, Walker T G, and Saffman M 2010 {\it Phys. Rev. Lett.} {\bf 104} 010503

\REF{[10]} Beterov I I and Saffman M 2015 {\it Phys. Rev. A} {\bf 92} 042710

\REF{[11]} Belyansky R, Young J T, Bienias P, Eldredge Z, Kaufman A M, Zoller P, and Gorshkov A V 2019 {\it Phys. Rev. Lett.} {\bf 123} 213603

\REF{[12]} Auger J M, Bergamini S, and Browne D E 2017 {\it Phys. Rev. A} {\bf 96} 052320

\REF{[13]} Weimer H, Muller M, Lesanovsky I, Zoller P, and Buchler H P 2010 {\it Nat. Phys.} {\bf 6} 382

\REF{[14]} Browaeys A and Lahaye T 2020 {\it Nat. Phys.} {\bf 16} 132

\REF{[15]} Liu L R, Hood J D, Yu Y, Zhang J T, Hutzler N R, Rosenband T, and Ni K K 2018 {\it Science} {\bf 360} 900

\REF{[16]} Liu L R, Hood J D, Yu Y, Zhang J T, Wang K, Lin Y W, Rosenband T, and Ni K K 2019 {\it Phys. Rev. X} {\bf 9} 021039

\REF{[17]} Zhang J T, Yu Y, Cairncross W B, Wang K, Picard L R B, Hood J D, Lin Y W, Hutson J M, and Ni K K 2020 {\it Phys. Rev. Lett.} {\bf 124} 253401.

\REF{[18]} He X D, Wang K P, Zhuang J, Xu P, Gao X, Guo R J, Sheng C, Liu M, Wang J, Li J M, Shlyapnikov G V, Zhan M S 2020 {\it Science} {\bf 370} 331

\REF{[19]} Sheng C, Hou J Y, He X D, Wang K P, Guo R J, Mamat B, Xu P, Lin M, Wang J and Zhan M S 2022 {\it Phys. Rev. Lett.} {\bf 128} 083202

\REF{[20]} Lee W, Kim H, and Ahn J 2017 {\it Phys. Rev. A} {\bf 95} 053424

\REF{[21]} Sheng C, Hou J, He X, Xu P, Wang K, Zhuang J, Li X, Liu M, Wang J, and Zhan M S 2021 {\it 	Phys. Rev. Res.} {\bf 3} 023008

\REF{[22]} Schymik K N, Lienhard V, Barredo D, Scholl P, Williams H, Browaeys A, and Lahaye T 2020 {\it Phys. Rev. A} {\bf 102} 063107

\REF{[23]} Barredo D, Léséleuc S, Lienhard V, Lahaye T and Browaeys A 2016 {\it Science} {\bf 354} 1021

\REF{[24]} Barredo D, Lienhard V, Léséleuc S, Lahaye T and Browaeys A 2018 {\it Nature} {\bf 561} 79

\REF{[25]} Lee W, Kim H, and Ahn J 2016 {\it Opt. Express} {\bf 24} 9816

\REF{[26]} Kumar A, Wu T Y, Giraldo F, and Weiss D S 2018 {\it Nature} {\bf 561} 83

\REF{[27]} Endres M, Bernien H, Keesling A, Levine H, Anschuetz E R, Krajenbrink A, Senko C, Vuletic V, Greiner M, and Lukin M D 2016 {\it Science} {\bf 354} 1024

\REF{[28]} Singh K, Anand S, Pocklington A, Kemp J T and Bernien H 2022 {\it Phys. Rev. X} {\bf 12} 011040

\REF{[29]} Zhang J T, Picard L R B, Carincross W B, Wang K, Yu Y, Fang F and Ni K K 2022 {\it Quantum Sci. Technol.} {\bf 7} 035006

}

\end{multicols}

\end{document}